\documentclass[pre,twocolumn,showpacs]{revtex4}
\usepackage{dcolumn}
\usepackage{graphicx}
\usepackage{bm}

\begin{document}  

\title{Elastic response of a nematic liquid crystal to an immersed nanowire}

\author{Christopher J. Smith} \email{csmith8@uwo.ca}
\author{Colin Denniston} \email{cdennist@uwo.ca}
\affiliation{Department of Applied Mathematics, The University of Western Ontario, London, Ontario, Canada N6A 5B8}

\date{\today}

\begin{abstract}
We study the immersion of a ferromagnetic nanowire within a nematic
liquid crystal using a lattice Boltzmann algorithm to solve the full
three-dimensional equations of hydrodynamics.  We present an algorithm for 
including a moving boundary, to simulate a nanowire, in a
lattice Boltzmann simulation.  The nematic imposes a torque on a
wire that increases linearly with the angle between the wire and the
equilibrium direction of the director field.  By rotation of these
nanowires, one can determine the elastic constants of the nematic.   
\end{abstract}

\pacs{61.30.Pq, 47.11.-j, 83.80.Xz}

%83.80.Xz 	Liquid crystals: nematic, cholesteric, smectic, discotic, etc.
%47.11.-j 	Computational methods in fluid dynamics
%61.30.Jf 	Defects in liquid crystals
%61.30.Pq 	Microconfined liquid crystals: droplets, cylinders, randomly confined liquid crystals, polymer dispersed liquid crystals, and porous systems
%61.46.Df 	Nanoparticles
%61.30.Dk 	Continuum models and theories of liquid crystal structure

\maketitle

%%%%%%%%%%%%%%%%%%%%%%%%%%%%%%%%%%%
\section{\label{s:Introduction}Introduction}
In a pioneering theoretical study in 1973 de Gennes and Brochard
looked at the effect of dispersing long rod-like particles in a liquid
crystal \cite{Broch_deGennes70}.  Despite interesting predictions on
ferronematics and ferrocholesterics, little experimental work was done
at the time.  Very recently, however, there has been a resurgence of
interest in this subject motivated by the advent of nanofabrication
techniques to make such particles
\cite{NKNM96,CL01,Tanase01,HFHY05,Gorby06}.  Early results are already
very encouraging.  For example, solutions of just $0.1-0.2$wt\% of
carbon nanotubes in E7 have shown a dramatic increase ($\sim 30^o$) in
the nematic-isotropic transition temperature  \cite{LP02,DGYK05}. 

Recent research on the inclusion of spherical particles in a nematic
have shown that unique interactions between particles which can lead
to self-assembly \cite{PSLW97,Loud_Poul01,KRZ06}.  There has also been
considerable interest in generalizing micro-rheological tools, such as
examining the fluctuations of microspheres to measure local viscosities
\cite{PCW97,KKPKCP03}.  However, there are complications in using
spheres for micro-rheology in an anisotropic fluid like a liquid crystal
that are not present in isotropic fluids.  First, boundary conditions
at the surface of the sphere typically induce defects either on or
close to the surface of the sphere \cite{PSLW97,PW98,MDGP99,GA00}.
These significantly modify the director field of the liquid crystal
thus making the sphere far from a passive probe.  Second, these
spheres are typically manipulated via laser traps
\cite{MSBOPNN04,YYY04,SKKPL05}. However, the electric fields required
to trap a sphere also strongly affect the alignment of the liquid
crystal \cite{SRBOPZM06} again making the system non-passive. 

\begin{figure}
\includegraphics{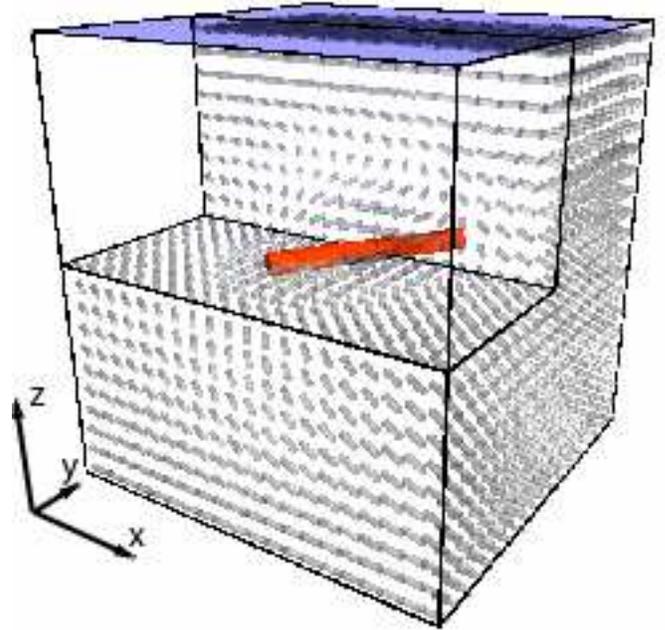}
\caption{Schematic of a sample cell in our simulation.  The dimensions
  are $1.25\mu m \times 1.25 \mu m \times 1.25 \mu m$.  Two rigid
  surfaces are located at $z=0$ and $z=L$ and there is periodic
  boundaries in the $xy-$plane.  In the middle of the cell is a
  ferromagnetic nanowire $625nm$ in length and approximately $88nm$ in
  diameter. The top front quarter has been removed for clarity.} 
\label{f:Setup}
\end{figure}

Recent experiments on nickel nanowires in a liquid crystal have shown
that they can be manipulated with magnetic fields and that the torque
exerted on the wire by the liquid crystal can be directly measured
\cite{LHSFRL04,LCRL05}. The magnetic fields required to manipulate
these ferromagnetic wires are typically very small, $\sim 5$ Gauss
\cite{LHSFRL04}, much less than that required to align the liquid
crystal ($\sim 3500$ Gauss for the Frederik's transition in 5CB in a
cell with thickness 10$\mu m$\cite{deGennes_Prost93}).  Further, when
the wire is aligned with the director field, it causes no
distortion. Thus, the use of such an anisotropic particle is ideal for
doing micro-rheology in an anisotropic fluid.  By manipulating the
wires immersed in the  nematic it should be possible to infer the
local viscous and elastic properties of the liquid crystal.

Simulations of colloids in liquid crystals have also recently been performed, but most have focused on homeotropic anchoring to the colloid.  Monte Carlo simulations were performed on rods of infinite length \cite{AASZ02}.  Also, there is work on soft two-dimensional cylinders using lattice Boltzmann \cite{LCH04} and FEM simulations on hard cylinders \cite{HGGAP06}.  All these works used homeotropic anchoring which leads to saturn ring-like defects along and around the cylinder.  In this paper we focus on the cylinder, or wire, with boundary conditions where the nematic aligns with the long axis of the wire.  The lack of defect formation from this configuration makes it ideal for microrheological applications as it can now measure the linear response of an unperturbed background.

Analytic calculations of the response of a liquid crystal to an
immersed particle can only be carried out by imposing rather unrealistic assumptions.  This makes the use of simulations necessary to understand precisely what is measured in experiments.  Lattice Boltzmann schemes have proved to be very successful in 
simulations of complex fluids \cite{Chen_Doolen98}, including liquid crystals \cite{Den_Or_Yeo00,Den_Or_Yeo01}.  We will examine a system similar to the experimental setup described in \cite{LHSFRL04}, as shown in Figure \ref{f:Setup}.  We have a simulation cell with rigid surfaces for boundaries at $z=0$ and $z=L$, where $L$ is the height of the cell.  The surfaces impose a preferred aligned direction for the nematic liquid crystal.  The $xy-$plane has periodic boundaries in both directions.  At the center of the cell, a small ferromagnetic nickel wire will be immersed.  
With the system in equilibrium, we can rotate the magnetic field in the $z$ plane to rotate the wire within the box.  Using the response function of the imposed torque, we are able to measure the elastic properties of the nematic liquid crystal. 

%%%%%%%%%%%%%%%%%%%%%%%%%%%%%%%%%%%

\section{\label{s:EquationsOfMotion}Equations of Motion}
Liquid crystals can be characterized by the second moment of the
orientation of the constituent molecules $\hat{p}$, in terms of a
local tensor order parameter ${\bf Q}$ where $Q_{\alpha \beta} =
\langle \hat{p}_\alpha \hat{p}_\beta - \frac{1}{3} \delta_{\alpha
  \beta} \rangle$.  We use angular brackets to denote a coarse-grained
thermal average.  Greek indices will be used to represent Cartesian
components of vectors and tensors and the usual summation over
repeated indices is assumed.  ${\bf Q}$ is a traceless symmetric
tensor.  We label its largest eigenvalue $\frac{2}{3}q$ so that  $0 <
q < 1$. $q$ describes the magnitude of order along {\bf Q}'s principle eigenvector $\hat {\bf n}$, referred to as the director.

The nematic order parameter evolves according to the convective-diffusive equation \cite{Beris_Edwards94}
\begin{equation} (\partial_t+{\vec u}\cdot{\bf \nabla}){\bf Q}-{\bf S}({\bf W},{\bf Q})= D_r {\bf H} 
\label{e:QEquation}
 \end{equation} 
where $D_r$ is a collective rotational diffusion constant.  The order parameter distribution can be both rotated and stretched by flow gradients described by 
\begin{eqnarray}
{\bf S}({\bf W},{\bf Q}) &=&(\chi{\bf A}+{\bf \Omega})({\bf Q}+{\bf I}/3)+({\bf Q}+ {\bf I}/3)(\chi{\bf A}-{\bf \Omega})\nonumber\\ & & -2\chi({\bf Q}+{\bf I}/3){\mbox{Tr}}({\bf A}({\bf Q}+{\bf I}/3))
\end{eqnarray}
where ${\bf A}=({\bf W}+{\bf W}^T)/2$ and ${\bf \Omega}=({\bf W}-{\bf
  W}^T)/2$ are the symmetric part and the anti-symmetric part
respectively of the velocity gradient tensor
$W_{\alpha\beta}=\partial_\beta u_\alpha$.  $\chi$ is a constant which
depends on the molecular details of a given liquid crystal\cite{footnote1}. 

The term on the right hand side of Eqn. (\ref{e:QEquation}) describes the relaxation of the order parameter towards the minimum of the free energy.  The molecular field ${\bf H}$ that provides the driving motion is related to the variational derivative of the free energy by 
\begin{equation}
{\bf H} = -\frac{\delta \mathcal{F}}{\delta {\bf Q}} + \frac{{\bf I}}{3} \textrm{Tr}\frac{\delta \mathcal{F}}{\delta {\bf Q}}.
\label{e:MolecularField}
\end{equation}
The symmetry and zero trace of ${\bf Q}$ and ${\bf H}$ are exploited
for simplification. 

The equilibrium properties of a liquid crystal can be described in 
terms of the Landau-de Gennes free energy \cite{deGennes_Prost93}  
\begin{equation}
\mathcal{F} = \int_V dV \left\{ F_B + F_E \right\} + \int_{\partial V} dS F_S.
\label{e:FreeEnergy}
\end{equation}
$F_B$ describes the bulk free energy 
\begin{eqnarray}
F_B & = & \frac {A_0}{2} (1 - \frac {\gamma} {3}) Q_{\alpha \beta}^2 - \frac {A_0 \gamma}{3} Q_{\alpha \beta}Q_{\beta \gamma}Q_{\gamma \alpha} \nonumber\\ & & + \frac {A_0 \gamma}{4} (Q_{\alpha \beta}^2)^2.
\label{e:BulkEnergy}
\end{eqnarray}
The bulk free energy describes a liquid crystal with a first order phase transition at $\gamma = 2.7$ from the isotropic to nematic phase \cite{Doi_Edwards88}.  When $\gamma >3$ the isotropic phase is completely unstable.  Unless noted,
we use $\gamma=3.2$ and $A_0=1.5 \times 10^5 pN/(\mu m)^2$ which leads to $q=0.555$ in the unstressed
equilibrium uniaxial case.  We also used higher values of $q$ and
found no qualitative difference.

The free energy also contains an elastic term,
\begin{eqnarray}
F_E = \frac{L_1}{2} (\partial_\alpha Q_{\beta \gamma})^2+
\frac{L_2}{2} (\partial_\alpha Q_{\alpha \gamma})(\partial_\beta Q_{\beta \gamma})+\nonumber\\
\frac{L_3}{2} Q_{\alpha \beta}(\partial_\alpha Q_{\gamma \epsilon})
(\partial_\beta Q_{\gamma \epsilon}),
\label{e:ElasticEnergy}
\end{eqnarray}
where the $L$'s are the material specific elastic constants.  It is straightforward to map $L$'s to the Frank elastic constants (the splay elastic constant $K_1$, the twist elastic constant $K_2$, and the bend elastic constant $K_3$)\cite{Beris_Edwards94}.  The one elastic constant approximation $K_1=K_2=K_3=K$ can be easily achieved by setting $L_1>0$, $L_2=L_3=0$ \cite{footnote2}.  
One advantage of simulations is the ease of switching between such an approximation commonly used in analytical work and the full equations necessary to compare to experiment.

At the {\it surfaces} of the system we assume a pinning potential 
\begin{equation}
F_S = \frac{1}{2} \alpha_s (Q_{\alpha \beta} - Q_{\alpha \beta}^o)^2,
\label{e:SurfaceEnergy}
\end{equation}
where ${\bf Q}^o$ has the form 
\begin{equation}
Q_{\alpha \beta}^o = q^o  \left( n^o_\alpha n^o_\beta - \frac{\delta_{\alpha \beta}}{3} \right),
\end{equation}
and $q^o$ is set to the equilibrium bulk value $q$.  This corresponds to a preferred direction ${\bf n}^o$ for the director at the surface.  Typically we set $\alpha_s = 6.33 \times 10^4 pN/\mu m$, which corresponds to the strong pinning limit where $Q \approx Q_o$ at the boundaries.

Inclusion of the wire in the system will be approximated as a moving
boundary.  We will use a pining potential similar to
(\ref{e:SurfaceEnergy}) for pining the director field of the liquid
crystal to the long axis of the wire.  This will be discussed in
greater detail below and in Section \ref{s:Nanowire}.  

The fluid momentum obeys the continuity
\begin{equation}
\partial_t \rho + \partial_{\alpha} \rho u_{\alpha} =0
\label{e:ConservationMass}
\end{equation}
and the Navier-Stokes equation
\begin{eqnarray}
& & \rho(\partial_t+ u_\beta \partial_\beta) u_\alpha = \partial_\beta \tau_{\alpha\beta}+\partial_\beta \sigma_{\alpha\beta}\nonumber\\ & & \quad+\eta \partial_\beta((1-3\partial_\rho P_{0}) \partial_\gamma u_\gamma\delta_{\alpha\beta}+\partial_\alpha u_\beta+\partial_\beta u_\alpha)
\label{e:NavierStokes}
\end{eqnarray}
where $\rho$ is the fluid density and $\eta=\rho \tau_f/3$ is an isotropic viscosity.  The form of this equation is not dissimilar to that for a simple fluid (anisotropic viscosities arise due to terms in $\sigma_{\alpha \beta}$. However the details of the stress tensor reflect the additional complications of liquid crystal hydrodynamics. There is a symmetric contribution
\begin{eqnarray}
\sigma_{\alpha\beta} &=&-P_0 \delta_{\alpha \beta} \nonumber\\
&-&\xi H_{\alpha\gamma}(Q_{\gamma\beta}+{1\over
  3}\delta_{\gamma\beta})-\xi (Q_{\alpha\gamma}+{1\over
  3}\delta_{\alpha\gamma})H_{\gamma\beta}\nonumber\\
&+& 2\xi
(Q_{\alpha\beta}+{1\over 3}\delta_{\alpha\beta})Q_{\gamma\epsilon}
H_{\gamma\epsilon}-\partial_\beta Q_{\gamma\nu} {\delta
{\cal F}\over \delta\partial_\alpha Q_{\gamma\nu}}
\label{e:SymmetricStress}
\end{eqnarray}
and an antisymmetric contribution
\begin{equation}
 \tau_{\alpha \beta} = Q_{\alpha \gamma} H_{\gamma \beta} -H_{\alpha
 \gamma}Q_{\gamma \beta} .
\label{e:AntisymmetricStress}
\end{equation}
These additional terms can be mapped onto the Erickson-Leslie equations to give the Leslie coefficients \cite{Den_Or_Yeo01}.  The isostatic pressure $P_o$ is constant in the simulations to a very good approximation, consistent with the liquid crystal being incompressible.

At the surface of the wire we assume a molecular surface field of a
form similar to Eqn (\ref{e:SurfaceEnergy}) where
\begin{equation}
H^w = w_s ({\bf Q} - {\bf Q}^w),
\label{Hwire}
\end{equation}
and $Q^w_{\alpha\beta}=q^w(m_\alpha m_\beta-\delta_{\alpha\beta}/3)$.
The director ${\bf m}$ of ${\bf Q}^w$ is fixed along the long
axis of the wire and $q^w$ is set to the equilibrium bulk value 
\begin{equation}
q = \frac{1}{4} \left( 1 - 3 \sqrt{1 - \frac{8}{3\gamma}} \right).
\end{equation}  

For the uniaxial case, where $Q_{\alpha \beta} = q  \left( n_\alpha
n_\beta - \frac{\delta_{\alpha \beta}}{3} \right)$, the torque on the
liquid crystal due to the presence of the wire is easily shown to be
\cite{deGennes_Prost93}
\begin{equation}
{\bm \Gamma} = {\bf m} \times {\bf h}^w,
\label{uniaxialtorque}
\end{equation}
where ${\bf m}$ is the unit direction vector of the wire, and 
\begin{equation}
h_\mu^w = q^w \left( m_\beta H_{\mu \beta}^w + m_\alpha H_{\alpha
  \mu}^w \right). 
\end{equation}
In our simulations, the torque is determined in system variables by 
\begin{eqnarray}
\Gamma_\alpha & = & 2 [Q_{\alpha\beta}H_{\alpha\gamma}^{w} -
  Q_{\alpha\gamma}H_{\alpha\beta}^{w} +
  H_{\beta\gamma}^{w}(Q_{\beta\beta}- Q_{\gamma\gamma}) + \nonumber \\
  & & \quad Q_{\beta\gamma}(H_{\beta\beta}^{w} -
  H_{\gamma\gamma}^{w})], 
\label{e:SystemTorque}
\end{eqnarray}
which can be easily shown to be equivalent to
Eqn. (\ref{uniaxialtorque}) in the uniaxial case.

The motion of the wire is performing a balancing act between
forces. Its motion is described by  
\begin{equation}
\Lambda (\hat{{\bf m}} \times \dot{\hat{{\bf m}}}) = - ({\bm \Gamma} - ({\bm \Gamma} \cdot \hat{{\bf m}}) \hat{{\bf m}}) + {\bm \mu} \times {\bf B}
\end{equation}
where $\Lambda$ is a rotational viscosity and ${\bm \mu} \times {\bf
  B}$ is the torque on the wire due it its magnetic moment ${\bm \mu}$
in the applied magnetic field.  The $(\Gamma\cdot {\bf m}){\bf m}$
term is zero but it's presence enhances the numerical stability of the
scheme.  In principle, $\Lambda$ should depend on the orientation of
the wire \cite{LCRL05}, but in this study we are interested in the
  steady state solution.  As such, this term only serves to move the
wire between steady states here.

%%%%%%%%%%%%%%%%%%%%%%%%%%%%%%%%%%%
\section{Numerical Implementation}

\subsection{\label{s:LatticeBoltzmannAlgorithm}The Lattice Boltzmann Algorithm for Liquid Crystal Hydrodynamics}
We summarize a lattice Boltzmann scheme which reproduces Eqns. (\ref{e:QEquation}), (\ref{e:ConservationMass}), and (\ref{e:NavierStokes}) to second order.  
The reader only interested in the results can safely skip over this section.
We defined two distribution functions $f_i({\bf x})$ and ${\bf G}_i({\bf x})$ where $i$ labels the lattice directions from site ${\bf x}$.  In two dimensions, we choose a nine-velocity model on a square lattice with velocity vectors $\overrightarrow{e}_i = (\pm 1, 0), (0,\pm 1), (\pm 1, \pm 1), (0,0)$.  For three-dimensional systems, a 15-velocity model on a cubic lattice with lattice vectors $\overrightarrow{e}_i = (0,0,0), (\pm 1,0,0), (0, \pm 1,0),(0,0,\pm1), (\pm 1, \pm 1, \pm 1)$ is chosen.  Physical variable are defined as moments of the distribution functions by 
\begin{equation}
\rho=\sum_i f_i, \qquad \rho u_\alpha = \sum_i f_i  e_{i\alpha},
\qquad {\bf Q} = \sum_i {\bf G}_i.
\label{e:Moments}
\end{equation}
The distribution functions evolve in a time step $\Delta t$ according to
\begin{eqnarray} 
&&f_i({\vec x}+{\vec e}_i \Delta t,t+\Delta t)-f_i({\vec x},t)=\nonumber\\ &&\quad\frac{\Delta t}{2} \left[{\cal C}_{fi}({\vec x},t,\left\{f_i \right\})+ {\cal C}_{fi}({\vec x}+{\vec e}_i \Delta t,t+\Delta t,\left\{f_i^*\right\})\right],\nonumber\\ 
\label{e:fEvolution}\\ &&{\bf G}_i({\vec x}+{\vec e}_i \Delta t,t+\Delta t)-{\bf G}_i({\vec x},t)= \nonumber\\ && \quad\frac{\Delta t}{2}\left[ {\cal C}_{{\bf G}i}({\vec x},t,\left\{{\bf G}_i \right\})+{\cal C}_{{\bf G}i}({\vec x}+{\vec e}_i \Delta t,t+\Delta t,\left\{{\bf G}_i^*\right\})\right].\nonumber\\ \label{e:GEvolution}
\end{eqnarray}
This represents free streaming with velocity ${\vec e}_i$ and a
collision step which allows the distribution to relax towards
equilibrium. $f_i^*$ and ${\bf G}_i^*$ are first order approximations
to $f_i({\vec x}+{\vec e}_i \Delta t,t+\Delta t)$ and ${\bf G}_i({\vec
  x}+{\vec e}_i \Delta t,t+\Delta t)$ respectively. They are obtained
by using only the collision operator ${\cal C}_{fi}({\vec
  x},t,\left\{f_i\right\})$ on the right of Equation
(\ref{e:fEvolution}) and a similar substitution in
(\ref{e:GEvolution}). Discretizing in this way, which is similar to a
predictor-corrector scheme, has the advantages that lattice viscosity
terms are eliminated to second order and that the stability of the
scheme is significantly improved.

The collision operators are taken to have the form of a single relaxation time Boltzmann equation\cite{Chen_Doolen98}, together with a forcing term
\begin{eqnarray}
{\cal C}_{fi}({\vec x},t,\left\{f_i \right\})&=&-\frac{1}{\tau_f}(f_i({\vec x},t)-f_i^{eq}({\vec x},t,\left\{f_i \right\}))\nonumber\\ & & \quad +p_i({\vec x},t,\left\{f_i \right\}), 
\label{e:fCollision}
\\ {\cal C}_{{\bf G}i}({\vec x},t,\left\{{\bf G}_i \right\})&=& -\frac{1}{\tau_{g}}({\bf G}_i({\vec x},t)-{\bf G}_i^{eq}({\vec x},t,\left\{{\bf G}_i \right\}))\nonumber\\ & & \quad+{\bf M}_i({\vec x},t,\left\{{\bf G}_i \right\}).
\label{e:GCollision}
\end{eqnarray}
The form of the equations of motion and thermodynamic equilibrium follow from the choice of the moments of the equilibrium distributions $f^{eq}_i$ and ${\bf G}^{eq}_i$ and the driving terms $p_i$ and ${\bf M}_i$. Full details of the algorithm can be found in \cite{Den_Or_Yeo01,Den_Mar_Or_Yeo04} for two-dimensional and three-dimensional simulations respectively.

%%%%%%%%%%%%%%%%%%%%%%%%%%%%%%%%%%%
\subsection{\label{s:Nanowire}Description of the Nanowire}
When a particle is suspended in a nematic liquid, anchoring boundary conditions at the particle's surface dictate the direction of the director field in the vicinity of the particle.  Using a pinning potential of the form of Eqn (\ref{e:SurfaceEnergy}), with $n_o$ along the long axis of the wire,  ensures the director of the nearby nematic is aligned to the nanowire.  The nanowire is ferromagnetic, enabling an external magnetic field to rotate the wire within the system.  

A numerical algorithm for simulating the inclusion of a nanowire in a liquid crystal system is presented in this section.  We could use an adaptive mesh with boundaries aligned on the wire but since the wire diameter is quite small
($\sim 100\,nm$), we decide against this method due to the small resolutions required.  
For instance, if there were 10 mesh points spanning the diameter of wire we would be approaching molecular resolution. 
The alternative is to pixelate a line onto a grid, a problem common in computer graphics.  First we will start off with a two-dimensional description, then expand to three dimensions.  We will describe both a standard Bresenham line algorithm \cite{Bresenham65}, expanded upon by Wu \cite{Wu91}, to produce lines on a two-dimensional surface and then we will present our current algorithm.

%%%%%%%%%%%%%%%%%%%%%%%%%%%%%%%%%%%
\subsubsection{\label{ss:WuLines}{\bf Computer Graphics Method}}

\begin{figure}
\includegraphics{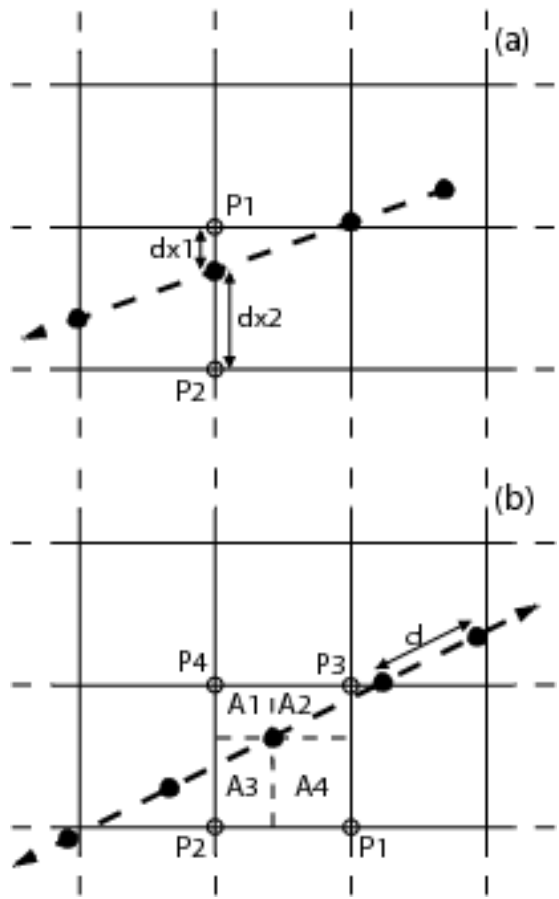}
\caption{(a) Example of the Brensenham anti-aliased line drawing algorithm.  If the slope is less one, the nodes are positioned along vertical lattice intersections, shown as the open dots at positions $P1$ and $P2$.  Value of the lattice nodes are inversely proportional to the distance from the lattice nodes to the wire nodes. (e.g. $P1=dx_2/dx$) For a slope greater than one, the horizontal lattice intersections are used.  (b) Sample cell of a two-dimensional system with an intersecting wire.  Along the wire, nodes are positioned at fixed distances $d$  apart to interact with the lattice nodes.  Each node is affected by the wire proportionally by the opposite enclosed area relative to the cell area. (e.g. $P2$ is affected from the wire by an amount $A2/dx^2$.)}
\label{f:WireNodes}
\end{figure}

The standard algorithm for pixelating a line is based on Bresenham's algorithm, generalized by Wu \cite{Bresenham65}.  This is illustrated in Fig \ref{f:WireNodes}(a).  When the slope of the line is less than 1, the points where the wire crosses vertical grid lines are marked.  The value of a particular wire node is proportional to the distance of the lattice crossing to the opposite node over the cell height.  For example, the value at $P2 = dx_1/dx$ in Figure \ref{f:WireNodes}(a).  The endpoints of the wire prove to be more cumbersome, involving extrapolation to the next grid line and weighting the points.  If the slope exceeds one, we swap the roles of $x$ and $y$ and exchange the vertical lattice crossings, with horizontal lattice crossings.  This method is easy to implement in two-dimensions. Unfortunately it is not easily generalized for three dimensional systems.  Other problems arise due to the fact that the number of nodes along the wire is not conserved as the wire rotates.  These issues will described in more detail in the next section.

%%%%%%%%%%%%%%%%%%%%%%%%%%%%%%%%%%%
\subsubsection{\label{ss:OurMethod}{\bf Node distribution method}}
Placing a wire within the simulation will require pixelation to map it
onto the lattice coordinates.  Our method is illustrated in Fig
\ref{f:WireNodes}(b).  We divide the wire into a fixed number of nodes
$N$.  The solid dots are the location of the nodes along the wire
separated by a distance $d$, where $d$ is incommensurate with the
lattice spacings.  Typically we have used $d=0.2 \phi_0\, dx$, where 
$\phi_0=0.618\cdots$ is the golden mean.  The wire nodes are then
distributed to the corners of the cell in which they reside.
To distribute the wire points to the closest lattice coordinates, the
ratio of the opposite area from the node over the cell area will be
calculated (Ex. $P1=A1/dx^2$).  It is easy to see that the node weight
thus defined varies smoothly from 0, when the wire node is at the
opposite corner of the cell, to 1, when the wire node goes through the
lattice point in question.  This is done for all nodes along the
nanowire.  If multiple nodes affect a single lattice coordinate, the
summation of their distributed values is used.  Although the
description above is for a two dimensional system, it is easily
generalized to three dimensions where ratios of opposite volumes to
the cell volume is used instead, and is the method employed in our simulations.

\begin{figure}
\includegraphics{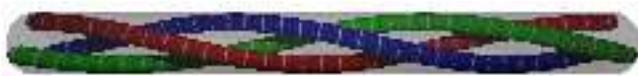}
\caption{Three chiral wires are shown together to simulate one larger wire.  The white spaces are sample locations of the nodes along the smaller wires.  To avoid commensurability with the lattice, the node distance is irrational.  Chirality of the wire is also included to avoid commensurability issues.  Each small wire is coiled by $2\pi$ about the central axis from end to end.}
\label{f:ThreeWires}
\end{figure}

In three dimensions an area of fluid will be displaced and affected by
the wire.  To give the wire a lateral dimension greater than $dx$, we
simulated many smaller wire strands in unison for the same effect.  A
diagram of three smaller strands forming a single wire is shown in
Figure \ref{f:ThreeWires}.  In our simulations we use five strands in
unison to form the nanowire.  The white bands represent where the
nodes lie.  Each strand, from end to end, is coiled about the central
axis by $2\pi$.  This twist on the strands helps make the wire nodes even more incommensurate with the underlying lattice, something that turns out to be highly desirable for producing a smooth response.

We want the surface energy of the wire to be
independent of the way it is discretized.  In other words, independent
of the number of strands and nodes used (assuming a sufficiently
high number) and independent of the orientation with respect to the
lattice (i.e. if we rotate the numerical mesh but keep all the fields
fixed).  This requires that the integral of $H^w$, from
Eqn. (\ref{Hwire}), over the surface of the wire be independent of the
orientation of the wire with respect to the underlying lattice.  
This will be the case if the discretization of  $\oint a_s
d\textrm{A}$ is a constant independent of orientation and wire
discretization, where $a_s$ is proportional to the surface area per
node.  To calculate $a_s$, we have  
\begin{equation}
\oint a_s d\textrm{A} = 2\pi r L = \sum_{i_w,i_n} a_s
\label{e:WireSurface}
\end{equation}
where $r$ and $L$ are the radius and length of the wire respectively,
$i_w$ is the index for the wire strand, and $i_n$ is the index for the
nodes on wire $i_w$.  As a result, $\sum_{i_w,i_n}$ the total number
of nodes in all the strands. Solving for $a_s$ gives
\begin{equation}
a_s = \frac{2 \pi r d}{N_s},
\end{equation}
where $d$ is the node spacing along the strands and $N_s$
is the number of strands used to create the wire.  

%%%%%%%%%%%%%%%%%%%%%%%%%%%%%%%%%%%
\section{Results}

\subsection{\label{s:Capacitance}{\bf Theory}}

\begin{figure}
\includegraphics{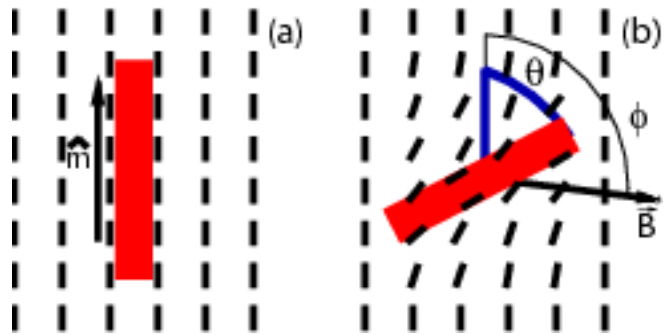}
\caption{(a) A schematic of the wire immersed in the nematic (black dashed) at equilibrium.  The magnetic field, the wire, and the director field of the nematic are all aligned along the same direction.  (b) Introducing a magnetic field at angle $\phi$, the wire reorients to an angle $\theta$ balancing the magnetic and elastic torques on the wire.  This leads to distortions in the nematic around the wire.}
\label{f:SimulationSetup}
\end{figure}

With no applied field, the long axis of the wire should line up parallel with the director, as shown in Figure \ref{f:SimulationSetup} (a).  Distortions in the orientation of the wire from this position, as in Figure \ref{f:SimulationSetup} (b) will cost elastic energy.  The director distorts to satisfy the boundary condition at the wire surface by introducing an elastic torque $\Gamma$ on the system.

In the equilibrium case where ${\bf Q}$ is uniaxial, $Q_{\alpha\beta}=\frac{3}{2}q(n_\alpha n_\beta - \frac{1}{3} \delta_{\alpha \beta})$, the elastic free energy density (\ref{e:ElasticEnergy}) can be rewritten as 
\begin{eqnarray}
F_{el} & = & \frac{1}{2} \left[ K_1 ({\bm \nabla}\cdot{\bf n})^2 + K_2 ({\bf n} \cdot {\bm \nabla} \times {\bf n})^2 \right. \nonumber \\ & & \left. + K_3 ({\bf n} \times {\bm \nabla} \times {\bf n})^2 \right]. \label{e:FrankFreeEnergy}
\end{eqnarray}
For analytic work it is useful to work in the approximation where the elastic constants are all
equal.  In that case, choosing the director field ${\bf n}$ to have the form $n_x = \cos \theta \sin \psi$, $n_y = \sin \theta \sin \psi$, and $n_z = \cos \psi$ then (\ref{e:FrankFreeEnergy}) can take the form 
\begin{eqnarray}
F_{el} & = & \frac{1}{2}K \left[ ({\bm \nabla} \psi)^2 + \sin^2 \psi ({\bm \nabla} \theta)^2 \right. \nonumber \\ & & +\left.2 \sin \psi \ {\bf n}\cdot  ({\bm \nabla}\theta \times {\bm \nabla} \psi) \right]. \label{e:AngleEnergy}
\end{eqnarray}
Minimizing (\ref{e:AngleEnergy}) with respect to the angles leads to two equations of equilibrium \cite{Chan92},
\begin{eqnarray}
\nabla^2 \psi - \sin \psi \cos \psi (\nabla \theta)^2 = 0 \label{e:OutOfPlane}\\
\sin \psi \nabla^2 \theta + 2 \cos \psi \nabla \psi \cdot \nabla \theta = 0. \label{e:InPlane}
\end{eqnarray}
In equilibrium, the nanowire will be positioned in the middle of the sample cell parallel to the top and bottom surfaces.  Using the coordinate system shown in Figure \ref{f:Setup}, $\psi = \frac{\pi}{2}$ in equilibrium.  If we assume that rotating the wire in the plane parallel to the walls produces no
out-of-plane deformations, then $\psi$ will remain constant at $\pi/2$.  Under this assumption  
Eqn. (\ref{e:OutOfPlane}) and (\ref{e:InPlane}) reduce to simply the Laplacian 
\begin{equation}
\nabla^2 \theta = 0. \label{e:Laplacian}
\end{equation}
If we further assume that the shape of the wire can be approximated as a prolate spheroid, the equations can be solved analytically.  The appropriate coordinate change is then
\begin{eqnarray}
r_\pm = \sqrt{\left(z \pm \frac{a}{2}\right)^2 + x^2 + y ^2} \label{e:rdefn} \\
\xi = \frac{r_+ + r_-}{a} \label{e:xidefn}\\
\eta = \frac{r_+ - r_-}{a} \\
\Xi = \tan^{-1}\left(\frac{y}{x}\right),
\end{eqnarray}
where $a$ is the interfocal distance.   Setting $\theta_o$ to a
constant equal to the wire orientation over the surface (i.e. ${\bf m}=(\cos\theta_o,\sin\theta_o,0)$), the Laplacian (\ref{e:Laplacian}) becomes easily solvable \cite{Morse53} with solution 
\begin{equation}
\theta = \theta_o \frac{\ln \left( \frac{\xi + 1}{\xi - 1} \right)}{\ln \left( \frac{\xi_o + 1}{\xi_o - 1} \right)},
\end{equation}
where the constant $\xi_o$ defines the prolate spheroidal surface.  From equilibrium, we integrate the elastic free energy density (\ref{e:AngleEnergy}) over all space, 
\begin{eqnarray}
U_{el} & = & \frac{K}{2} \int dV (\nabla \theta)^2 \nonumber \\
& = & 2 \pi a K \theta_o^2 \left[ \ln \frac{\xi_o + 1}{\xi_o -1} \right]^{-1}, \label{e:ProlatePotential}
\end{eqnarray}
which results in an expression for the potential of a prolate spheroidal inclusion in a nematic.  

In the approximation that the three elastic constants are equal, $K_1 = K_2 = K_3 = K$, Brochard and de Gennes \cite{Broch_deGennes70} calculated the energy involved in considering a long inclusion, such as a wire, within a nematic as a function of the inclusions orientation.  In the case the anchoring at the surface was along the long axis of the wire, they exploited an analogy from electrostatics of an object at fixed potentials to predict that the energy varies with the angle of orientation as 
\begin{equation}
U_{el} = 2\pi CK \theta^2,
\label{e:TorqueToC}
\end{equation}
where $C$ is the capacitance of the wire.  Equating (\ref{e:ProlatePotential}) and (\ref{e:TorqueToC}), and making use of Eqn. (\ref{e:rdefn}) and (\ref{e:xidefn}), the capacitance is 
\begin{eqnarray}
C & = & a \left[ \ln \frac{\xi_o + 1}{\xi_o -1} \right]^{-1} \\
& = & 2 L \sqrt{1- \left( \frac{R}{L} \right)^2} \left[ \ln \left( \frac{1 + \sqrt{1- \left( \frac{R}{L} \right)^2}}{1 - \sqrt{1- \left( \frac{R}{L} \right)^2}} \right) \right]^{-1}, \label{e:RealCapacitance}
\end{eqnarray}
where $R$ and $L$ are the radius and length of the wire respectively.  In the limit $\frac{R}{L} \rightarrow 1$ the capacitance is simply $C \rightarrow R$, that of a sphere, as expected.  As $\frac{R}{L} \rightarrow 0$, the capacitance is 
\begin{equation}
C \approx \frac{L}{\ln \left( \frac{2 L}{R} \right)}, \label{e:Capacitance}
\end{equation}
which is {\it not} the result Brochard and deGennes had attained,
their result containing two incorrect factors of 2.

Although Eqn. (\ref{e:Capacitance}) is the capacitance for a prolate ellipsoid, the wire in our system has a shape of a right cylinder.  It is not possible to solve Laplace's equation exactly for this shape, but expanding in 
$1/\zeta = 1/\ln (\frac{2 L}{R})$, Jackson\cite{Jackson00} derived an approximate expression, 
\begin{eqnarray}
C & = & L \left\{ \frac{1}{\zeta} + \frac{1}{\zeta^2} (1 - \ln 2) \right. \nonumber \\ & & + \left. \frac{1}{\zeta^3} \left[ 1 + (1-\ln 2)^2 - \frac{\pi^2}{12} \right] + O(1/\zeta^4) \right\}. \label{e:CorrectedCap}
\end{eqnarray}
For high aspect ratio wires, both Eqn. (\ref{e:Capacitance}) and (\ref{e:CorrectedCap}) are equivalent.  As we shall see below, for low aspect ratio wires there is significant differences between the two equations and Eqn. (\ref{e:CorrectedCap}) should be used instead.

%%%%%%%%%%%%%%%%%%%%%%%%%%%%%%%%%%%%%%%%%%%%%%%%%%%%

\subsection{\label{s:Something}{\bf Torque on a nanowire in the simulations}}

In our simulations we will first try to replicate as closely as
possible the theoretical results of the previous subsection.  To do
this we will first work in the one elastic constant limit and then
move on to more realistic situations.  This will demonstrate the
ability of the node distribution algorithm to reproduce both the
analytic results and the results from experiments.  

We start with a wire aligned with the nematic director field in
equilibrium,  as shown in Figure \ref{f:SimulationSetup}(a).  A
magnetic field aligned in the direction of the long axis of the wire
is introduced and slowly rotated in the $xy-$plane (See the schematic
in Fig.~\ref{f:SimulationSetup}(b) and the 3D simulation in
Fig.~\ref{f:Setup} for reference).  Taking $L_2=L_3=0$, to work in the
one-elastic constant approximation, and {\it if} there is no out-of-plane distortion, a measurement of the torque on the wire from the nematic as a function of rotation angle should yield the elastic constant, assuming the capacitance of the wire is known.  Using Eqn (\ref{e:TorqueToC}) for the potential energy results in a torque on a wire not aligned with the nematic,  
\begin{equation} 
\Gamma = \frac{\partial U}{\partial \theta} = 4 \pi C K \theta.
\end{equation} 
Eliminating the dependence on $\theta$, one can simply measure the value of the elastic constant via, 
\begin{equation}\
\frac{\partial \Gamma}{\partial \theta} = 4\pi K C.
\label{e:TorqueSlope}
\end{equation}
Thus, if $K$ is known one can determine $C$ or {\it vice versa}.

\begin{figure}
\includegraphics{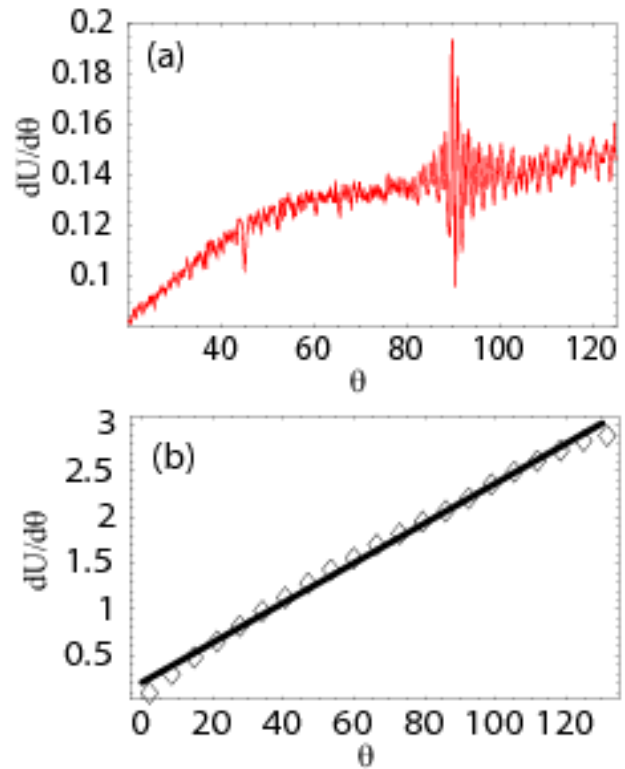}
\caption{(a)Wu's algorithm only works in 2-dimensions.  The system is $1.25\mu m \times 1.25\mu m$ with a plate of length $0.675 \mu m$.  The torque clearly exhibits commensurabilty problems along certain orientations of the wire. (b) Using the algorithm described in Section \ref{ss:OurMethod}, one notices a smooth torque along a rotation.  The are no indications of any commensurabilty problems as in Wu's algorithm.  The system is $1.25\mu m \times 1.25\mu m \times 1.25\mu m$ with a wire of length $0.675 \mu m$.  The one elastic constant approximation is used with $K=10.72pN$.  To measure the slope a least squares fit is performed from the 0 to $\sim$2 radians.
}
\label{f:TorqueComparisons}
\end{figure}

Using the two methods described in Sections \ref{ss:WuLines} and
\ref{ss:OurMethod}, we can simulate a wire in our system and measure
the torque imposed by the nematic.  The results are shown in
Fig.~\ref{f:TorqueComparisons}.  It should be immediately apparent
that using the algorithm borrowed from the graphics industry, Wu's
anti-aliased line method, produces poor results
(Fig.~\ref{f:TorqueComparisons} (a)).  Aside from being rapidly
fluctuating, there are profound commensurability effects when the wire
is at $\pi/4$ and $\pi/2$. The node distribution algorithm described
in Section \ref{ss:OurMethod} produces a smooth linear relationship
between $\Gamma_z$ and $\theta$, as expected in the one-elastic
constant approximation \cite{LHSFRL04}, as shown in Fig
\ref{f:TorqueComparisons}(b).  As mentioned in Sections
\ref{ss:WuLines}, Wu's method does not conserve the number of nodes
along the wire (i.e. the wire ``density'' is not constant).
Therefore, as the wire rotates the torque, which involves an integral
over the surface of the wire, fluctuates.  In contrast, the node
distribution algorithm we introduced, conserves the wire ``density''
at all times, as described by Eqn (\ref{e:WireSurface}).  The number
of nodes along the wire is constant for the algorithm we proposed,
resulting in almost no observable effects due to the orientation of the
wire with respect to the underlying lattice.  This leads to a smooth
dependence of the torque on the angle of the wire with respect to the
background nematic.  Therefore, only the node distribution algorithm
described in Section \ref{ss:OurMethod} will be used for
subsequent simulations. 

\begin{figure}
\includegraphics{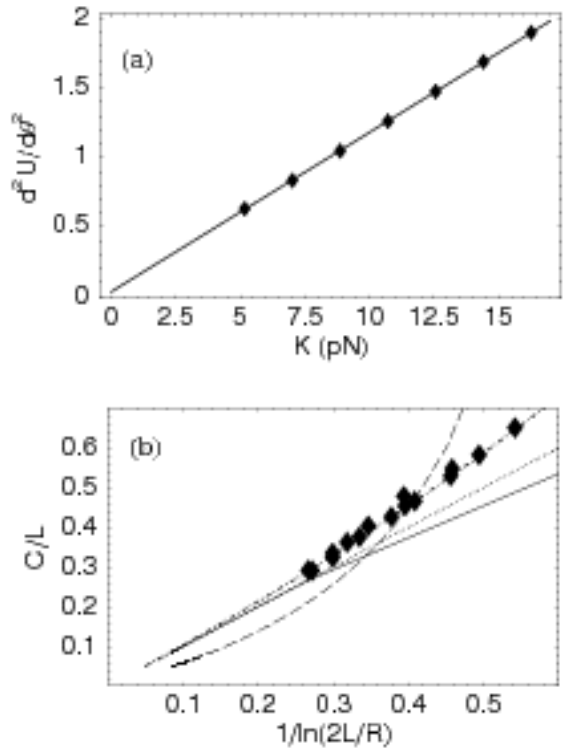}
\caption{(a) The slope of the torque of a 625$nm$ wire as a function of the elastic constant.  The one-elastic constant approximation is used in these simulations.  (b) The relationship of capacitance as a function of lengths and radii are explored.  Lengths vary from 625$nm$ to 1.25$\mu m$ and radii from 44$nm$ to 198$nm$.  The diamonds are the results from our simulations.  The exact solution Eq. (\ref{e:RealCapacitance}) and the first order approximation, Eq. (\ref{e:Capacitance}) for the capacitance of a prolate spheroid are the solid and dotted lines respectively.  The dashed line is the solution of Brochard and deGennes \cite{Broch_deGennes70}.  The dash-dot line which traverses through the data is the third order approximation, Eqn. (\ref{e:CorrectedCap}), for the capacitance of a prolate spheroid from Jackson.}
\label{f:KVariations}
\end{figure}

We are now in a position to test Eqn. (\ref{e:TorqueSlope}).
Within the single elastic constant approximation the relationship
between the capacitance of the wire and the elastic constant of the
nematic is approximately linear, as suggested by Eqn
(\ref{e:TorqueSlope}).  Using a wire of length $625nm$ and radius
$88nm$ immersed within a nematic in a box of dimensions $1.25\mu m
\times 1.25\mu m \times 1.25 \mu m$,  the elastic constant of the
nematic $K$ was varied from $5.2pN$ to $16.3pN$ in $1.8pN$ increments.
For each elastic constant, the wire is rotated from $0$ to $\pi$ and
the slope of the torque versus $\theta$ curve is measured (e.g. the
slope of the solid line in Fig.~\ref{f:TorqueComparisons}(b) provides
one of the points in Fig.~\ref{f:KVariations}(a)).   The resulting points
are plotted in  Figure \ref{f:KVariations} (a).  As expected, we see a
linear relationship (the solid line) between $K$ and the measured slope of the 
torque $\Gamma$ (diamonds) as expected.  The slope of this line yields
$4\pi C$, as can be seen from Eqn. (\ref{e:TorqueSlope}).

Having verified the linear relationship in Eqn. (\ref{e:TorqueSlope}),
we can now turn the situation around and use the known value of $K$
put into the simulation to measure the unknown $C$.  This quantity is
not experimentally assessable unless an independent measure of the
elastic constant $K$ is available.
We simulated 15 systems where the
length of the wire was varied from $625nm$ to $1.875\mu m$ and the
diameter from $22nm$ to $197nm$.  This resulted in a minimum aspect
ratio of $\sim 3$ to a maximum of $\sim 42$ for
the nanowires in our simulations.  To avoid finite size effects, we
kept the simulation box larger than twice the length of the wire
(boxes larger than this gave comparable results).  The one elastic
constant approximation was used for all simulations, with $K=10.72 pN$.  
Figure \ref{f:KVariations} (b) shows our results for $C/L$ as a
function of $1/\ln(2L/R)$.  The dashed line is the incorrect
prediction from Brochard and de Gennes which clearly does not predict
the capacitance of a prolate spheroid in a correct manner.  In Section
\ref{s:Capacitance}  we derived the correct expression for the
capacitance of a prolate spheroid, Eqn. (\ref{e:RealCapacitance}), as
represented by the solid line.  The first order approximation is shown
by the dotted line.  Both of these forms for the capacitance converge
for high aspect ratio wires, but fail for low aspect ratio wires.
Interestingly, the first order approximation is nearer the simulation
results for the wires than the exact solution for the capacitance of a
prolate spheroid.  For low aspect ratio wires, the expansion of
Jackson for a right circular cylinder must be used, and one can see
close agreement between this third order approximation and our
results.  All three theoretical lines
(Eqn.(\ref{e:RealCapacitance})-(\ref{e:CorrectedCap})) converge for
high aspect ratio wires.  For aspect ratios less than $\sim 75$ Eqn. (\ref{e:CorrectedCap}) separates from both Eqn. (\ref{e:RealCapacitance}) and
Eqn. (\ref{e:Capacitance}) and is obviously the correct expression to
use.  If one were to use the wires as
micro-rheological probes, they would probably need to be a short,
low aspect ratio wire (to fit in an {\it in-vivo} environment) and similar
to the sizes studied within this paper. 

\begin{figure}
\includegraphics{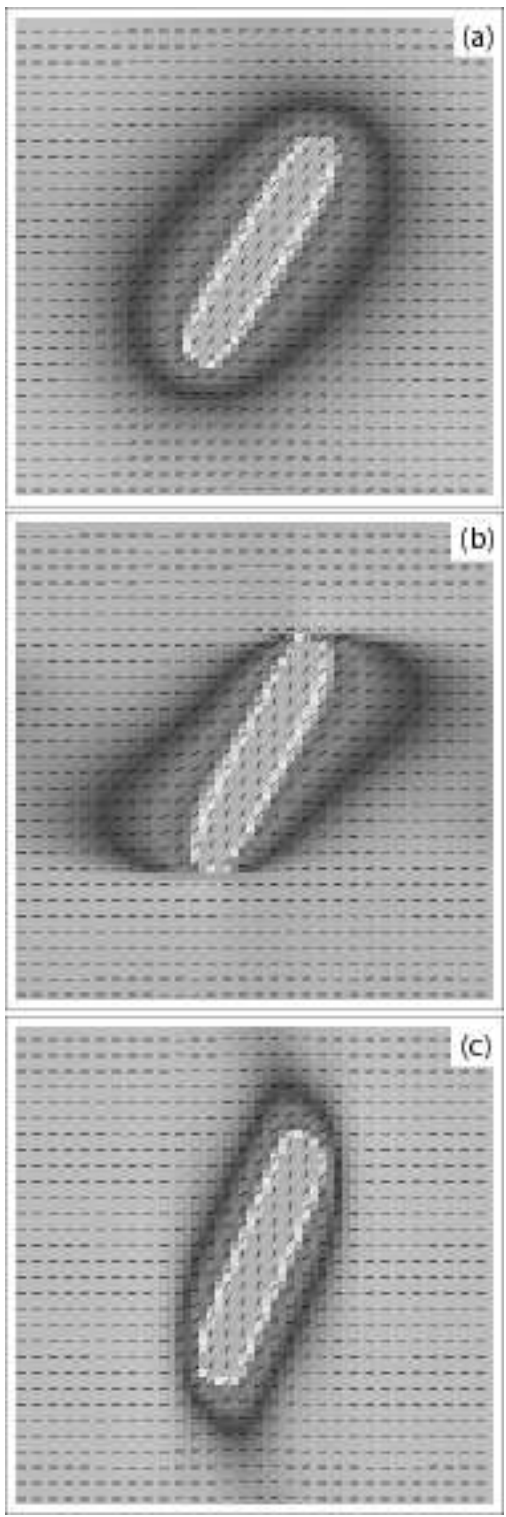}
\caption{Cross sections of three systems of size $3.75\mu m \times
  3.75\mu m \times 3.75 \mu m$ with a wire of length $1.875 \mu m$
  immersed in the center.  The wire was rotated from equilibrium to an
  angle of $\frac{3\pi}{8}$.  The shading indicates the inplane
  nematic angle with the director field of the nematic superimposed.
  (a) The one elastic constant approximation where $K_1=K_2=K_3$.  The
  director field of the nematic is approximately isotropic in the
  region surrounding the wire.  In (b) $K_1 < K_3$ and in (c) $K_1 > K_3$}
\label{f:Comparison}
\end{figure}

In real liquid crystals the elastic constants are {\it not} equal
(e.g. for 5CB, $K_1$ and $K_3$ differ by a factor of $\sim 1.6$
\cite{CTF89}).  However, the torque is still observed to be a linear
function of $\theta$ in experiments \cite{LHSFRL04,LCRL05}.  This
brings up the question of which elastic constant, or combination
of elastic constants, is measured in these experiments.
As the wire rotates through the nematic, one of the modes of
deformation, splaying, twisting, or bending, should be favorable for
the nematic.  The distortion field of the nematic will be greatly
affected by the relative differences of the strengths in the various
elastic constants.   Figure \ref{f:Comparison} is a density plot of
the nematic angle with the director field of the nematic superimposed
for three systems with varying elastic constants (A cross section
through the center of the cell is shown).  In all three panels, the
wire was immersed in the center of the nematic as in figure
\ref{f:Setup}, and rotated $\frac{3\pi}{8}$ from equilibrium.  In (a)
the one elastic constant approximation is used with $K=10.72pN$.  The
nematic angle forms homeotropic coronas in the region immediately
surrounding the wire.  In (b) and (c), the one elastic constant
approximation is no longer valid as various elastic constants are
used.  For (b), $K_1 = 15.34pN$, $K_2 = 3.88pN$, and $K_3 = 8.5pN$ and
in (c) the values of $K_1$ and $K_3$ are interchanged.  Looking at the
density plot of the inplane angle, it is discernible there are clear
differences in the distortion pattern the nematic uses to minimize its
free energy for these two systems.  The splay elastic constant is
larger in (b), which explains the distortion field of the nematic
exhibiting a higher tendency to bend within the system.  Interchanging
the elastic constants of splay and bend, as in (c), the distortion
field is dominated by splay, as expected.

\begin{figure}
\includegraphics{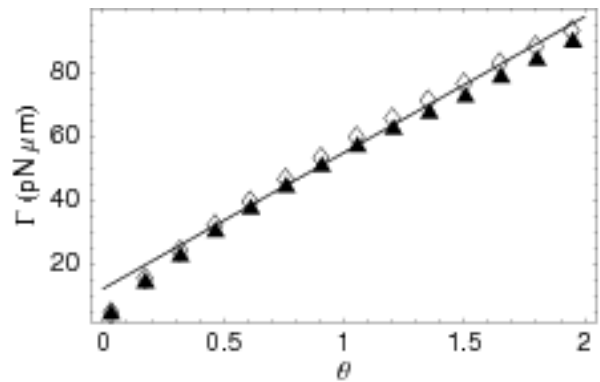}
\caption{Comparison of the torques of two systems of size $3.75\mu m \times 3.75\mu m \times 3.75 \mu m$ with a wire of length $1.875 \mu m$.  The elastic constants of torque represented by the open diamonds are $K_1 = 15.34pN$, $K_2 = 3.88pN$, and $K_3 = 8.5pN$ and of the filled triangles are the same elastic constants aforementioned with $K_1$ and $K_3$ interchanged.  The solid line is a least squares fit to the data.}
\label{f:MultipleK}
\end{figure}

Observations of the in-plane nematic angle show discernible
differences for the two systems shown Figure \ref{f:Comparison} (b)
and (c).  However, the torques of the wires rotating through the two systems
are nearly identical, as shown in Figure \ref{f:MultipleK}.  Even
though the elastic constants of splay and bend are interchanged, the
torques are approximately equal implying that for
Eqn. (\ref{e:TorqueSlope}), the same value of $K$ is being measured.
Using a least squares fit to the data, the solid line in Figure
\ref{f:MultipleK}, the slope is $\sim 8 pN$ which is approximately the
value of the smaller of either splay or bend depending on the system
being observed.  Thus, experiments will measure the smaller of the
splay or bend elastic constant and not some average of the elastic
constants as was supposed in Ref.~\cite{LCRL05}.  Of course, it may
not be known {\it a priori} in an experiment which elastic constant is
smaller.  However, it should be straightforward to observe the
birefringence pattern through crossed polarizers and ascertain whether
it is more similar to the distortion field shown
Fig.~\ref{f:Comparison}(b) or (c) and thereby ascertain whether the
bend or splay dominates.

%%%%%%%%%%%%%%%%%%%%%%%%%%%%%%%%%%%
\section{\label{s:Conclusion}Conclusions}

In this paper we present an algorithm to model a moving inclusion in a nematic liquid crystal system.  From a knowledge of the capacitance of the ferromagnetic nanowire, it is possible to deduce elastic constants of the liquid crystal from a measurement of the torque.  Previously quoted expressions for the capacitance of the wire \cite{Broch_deGennes70} where found to be incorrect and the correct expressions, both for a prolate spheroid and for a right circular cylinder, were given.  Most wires used to date have had a very high aspect ratio ($\sim100$)
\cite{LCRL05}.  However, if one were to use the wires as an {\it in-vivo} probe of the local elastic properties in, say, a biological environment then much lower aspect wires, or prolate spheroids may be more appropriate.  In that case, (i.e.
for aspect ratios less than around 75) either the exact expression for the prolate spheroid or the third order expansion by Jackson for the appropriate capacitance should be used.

For realistic systems with multiple elastic constants, we found that the system will tend to trade between splay or bend distortions depending on which one cost less free energy.  As a result, the linear elastic response of the nematic to a rotation of the wire is dominated by the smaller of the splay or bend elastic constants.  Thus, experiments will typically measure the smaller of $K_1$ or $K_3$ and not some combination.  It should be possible to deduce the elastic constant being measured based on the distortion field of the nematic around the rotated wire, as demonstrated in Fig.~\ref{f:Comparison}.

In future works, we will examine the nonlinear response that occurs at wire rotations close to $\pi$ and the dynamical response. 

\acknowledgements
We would like to thank the Natural Science and Engineering Research Council of Canada (NSERC) and the Shared Hierarchical Academic Research Computing Network (SHARCNet) for financially supporting this study.

%\bibliography{/home/csmith/Documents/Thesis/PhDBibFile.bib}
%\bibliography{/home/grads/csmith/Colin/Paper/PhDBibFile.bib}

\end{document}